\documentstyle[11pt,paspconf]{article}

\begin{document}

\title{The FIRBACK 175$\mu$m Survey}
\author{David L. Clements, J-L Puget, G. Lagache, H. Dole, P. Cox, R. Gispert}
\affil{Institut d'Astrophysique Spatiale, Univ. Paris XI, F-91405 ORSAY Cedex,
France}
\author{W.T. Reach}
\affil{IPAC, Caltech, MS 100-22, Pasadena, CA 91125, USA}
\author{C. Cesarsky, D. Elbaz, H. Aussel}
\affil{Service d'Astrophysique, Orme des Merisiers, Bat. 709, CEA/Saclay,
F91191 Gif-sur-Yvette CEDEX, France}
\author{F. Bouchet, B. Guiderdoni, A. Omont}
\affil{IAP, Boulevard d'Arago, Paris}
\author{F-X. Desert}
\affil{Laboratoire d'Astrophysique, Observatoire de Grenoble, BP 53, 
414 rue de la piscine, 38041 Grenoble CEDEX 9,  France}
\author{A. Franceschini}
\affil{Dipartimento di Astronomia, Universita' di Padova, Padova, Italy}
\author{A. Moorwood}
\affil{European Southern Observatory, Garching, Germany}
\author{D. Lemke}
\affil{Max Planck Instiute for Astronomy, K\"{o}nigstuhl, Heidelberg, Germany}

\begin{abstract}
The FIRBACK project is a collaboration of eight institutes aimed at
using long wavelength surveys with ISO to uncover the nature of the
Cosmic Far-Infrared Background. We have surveyed nearly 4 sq. deg. in
three regions in the northern and southern hemispheres with the
ISOPHOT instrument on ISO, using the 175$\mu$m filter. As reported by
Puget et al. (1998) and Reach et al. (these proceedings),
the first 0.25 sq. deg. region studied in this programme contains many
more 175$\mu$m sources than expected.  Initial reduction of the
remaining survey areas appears to confirm this result.  The area
surveyed by FIRBACK represents the largest and deepest long wavelength
far-IR survey available before the launch of SIRTF.
\end{abstract}


\section{Introduction}
There has been considerable suspicion for some time that dust
obscuration may bias the optical view of the early universe. UV and
optical radiation from dusty objects would be reprocessed into thermal
emission from dust, and shifted from the optical into the far-IR, and
thus inaccessible from the ground. This effect becomes more
significant at higher redshifts where the observed-frame optical
corresponds to the near and far-UV in the emitted-frame. The IRAS
all-sky survey unveiled the local population of infrared galaxies
ranging from normal spirals to the spectacular luminous and
ultraluminous infrared galaxies, but only a handful of these sources
had a significant redshift. Our knowledge of the far-IR 'dark-side' of
galaxy formation is thus limited to the local universe.  We do,
however, know that there are individual objects at high redshift that
contain substantial amounts of dust. The gravitationally lensed IRAS
source 10214+4724 (Rowan-Robinson et al 1991, Serjeant et al 1995) is
perhaps the best-known high redshift IRAS galaxy, but dust has also
been detected in submillimetre observations of several quasars and
radio galaxies (Hughes et al 1997 and references therein). Information
on the dust properties of the bulk of galaxies at high redshift has so
far relied on measurements of the integrated cosmic far-IR background
(CIRB). Clear detection of this background has now been achieved
(eg. Guiderdoni et al 1997) using data from the COBE satellite. The
level of this background is substantially higher than was expected on
the basis of models that do not include a substantial amount of
dust-obscured star formation.  Models of the CIRB (Guiderdoni et
al. 1997) suggest that a significant fraction ($\sim$ 10\%) of the
objects contributing to the CIRB might be detectable by the ISO
satellite by observations using its longest wavelength cosmological
filter - the PHOT 175$\mu$m channel. The FIRBACK survey was devised to
make these observations. A separate survey in a small area at these
wavelengths by a Japanese team (Kawara et al. 1998) has also been
attempted. Their results are compatible with those of FIRBACK.

\section{The FIRBACK Survey}

The FIRBACK Survey was conducted in two regions selected for low
cirrus contamination and low galactic foreground. The two areas
selected were the Marano Field (Marano et al. 1988) and the ELAIS N1
survey region (Oliver et al. in preparation). Roughly one sq. deg.
was observed in the Marano Field, while about two sq. deg.  were
observed in ELAIS N1.  Additionally, a further $\sim$one sq. deg.  was
observed in the ELAIS N2 region in collaboration with the ELAIS survey
programme. The final survey area will thus be about 4 sq. dgrees. The
rest of this paper will concentrate only on the Marano and ELAIS N1
surveys since we have so far not processed the ELAIS N2 data.  All the
FIRBACK surveys were conducted with the PHOT instrument aboard ISO
using the C200 2x2 array detector and the C160 (175$\mu$m effective
wavelength) filter. Three different strategies were used for the
observations. The first Marano Field area covered, a 0.25
sq. deg. region dubbed Marano 1, is discussed extensively by
W.T. Reach in the present volume and Puget et al. (1998).  Four 19x19
rasters were obtained in this area. Each raster scan stepped by one
pixel in both scan and cross-scan directions so that each point on the
sky was observed by each of the four C200 pixels for an integration
time of 16s. This sequence was then repeated four times with an offset
of 92'', the size of one pixel.  We thus obtain either four completely
independent maps each with a 64s integration time, or a combined map
with a 256s integration time. See Reach et al. and Puget et al. (1998)
for further discussion.  For the remaining 0.75 sq. deg. in the Marano
region, termed Marano 2 to 4, we used a similar technique, except that
the four independent rasters are offset by $\sim$ 46'' (ie. half a
pixel) rather than 92''. This yields significantly better sampling of
the PSF. The ELAIS N1 survey uses a similar idea, but uses just two
independent rasters offset by 46''. We expect to reach detection
limits of $\sim$100mJy in all survey regions. Based on number counts
obtained in Marano 1, this should provide over 200 175$\mu$m sources
in the final 4 sq. deg. studied.  Other ISO data is also available for
some of these areas. The ELAIS N1 region has been surveyed by the
ELAIS team at 15 and 90 $\mu$m, while the Marano 1 area coincides with
the ISOCAM DEEP and ULTRADEEP surveys at 7 and 15 $\mu$m. The Marano 2
and 3 regions ($\sim$ 0.5 sq. deg.) were also surveyed at 15$\mu$m
during the course of the FIRBACK observations, so that only one 0.25
sq. deg. area, Marano 4, has no corresponding survey in the mid-IR.
We use a combination of standard PIA techniques and custom-written
software to reduce the FIRBACK data and to extract sources and
fluxes. To date only the initial Marano 1 area has been fully reduced
and had a final catalogue extracted. Preliminary analysis of the rest
of the FIRBACK data however supports the initial conclusions in Reach
et al. and Puget et al. (1998), namely that we find many more
175$\mu$m sources than would be expected on the basis of what we
currently know of the local infrared galaxy population.

\section{The Next Steps}

Once the ELAIS N1 and Marano 2--4 regions are fully reduced, we must
obtain identifications for our far-IR sources and obtain redshift
spectra for them.  This will not be an easy process since the PHOT
angular resolution is very poor. The locations of our 175$\mu$m
sources will be uncertain by over an arcminute, especially in the
Marano 1 region where the PSF is poorly sampled. We must thus rely on
observations at other wavelengths to provide identifications. The two
most useful wavelengths for this are likely to be the mid-IR, which is
why most of the FIRBACK survey regions have also been studied at
15$\mu$m, and the radio.
The mid-IR data will be able to detect hot dust associated with the same
processes that power the far-IR luminosity. In a typical nearby starburst
the mid-to-far-IR flux ratio is $\sim$4\%, though with a large
scatter, suggesting that a typical 100mJy far-IR source should have a
15$\mu$m flux of a few mJy. However, there are substantial variations
in the mid-to-far-IR ratio from object-to-object, while the complex
spectral features in the mid-IR mean that this ratio will strongly
vary with redshift. Nevertheless initial results suggest that a
substantial number of Marano 1 175$\mu$m sources have 15$\mu$m
counterparts. The ISOCAM data will provide source positions to $\sim$
6'' accuracy.
The well-known radio-far-IR relation (Helou et al. 1987) implies a
radio to 175 $\mu$m flux ratio of about 0.5\%, though with a large
scatter. Several hundred mJy 175$\mu$m sources might thus be expected
to correspond to radio sources of a few hundred $\mu$Jy. These fluxes
are readily detectable in long integrations with modern radio
interferometers and are thus be capable of identifying
the 175$\mu$m survey objects. Much of the ELAIS N1 area has been
surveyed by the ELAIS team using the VLA, while we have conducted ATCA
observations of the Marano regions. This will provide source positions with
an accuracy of $\sim$1'' or better.
With better positions from radio or mid-IR associations, we will be
able to determine an optical identification and then obtain a
redshift. For objects lacking a mid-IR or radio association, obtaining
an optical identification will be much more difficult and we may have
to rely on optical and infrared multicolour techniques to select
likely targets. For example, the FIRBACK sources are likely to contain
substantial masses of dust and may thus be redder than more normal
galaxies. Red objects within the positional error circle for a FIRBACK
source will thus be strong candidate identifications. The results of
the radio and mid-IR identifications will guide the techniques applied
at other wavelengths.

\section{Conclusions}

The FIRBACK 175$\mu$m survey will provide the first information on the
nature and origin of the newly-discovered cosmic far-IR background. The
early indications are that the background is the integrated emission of
a large number of point sources, of which FIRBACK is detecting about 10\%.
We must await the completion of the followup programme before the nature of
these sources is understood.
As far as surveys with the SIRTF instruments are concerned, FIRBACK is
the deepest far-IR large-area survey that will be completed before the
launch of SIRTF and provides a useful preview of what the 160$\mu$m
channel of MIPS will be capable of. With sensitive instruments, better
PSF sampling and a larger mirror, MIPS and SIRTF may well be able to
resolve most of the objects contributing to the far-IR
background. Once this has been achieved, we will have determined the
role of dust in the early universe, and have unveiled the
``dark-side'' of galaxy formation.
\acknowledgments
DLC is supported by the EC TMR Network programme, FMRX-CT96-0068. This
work is based on observations with ISO, an ESA project with
instruments funded by ESA Member States (especially the PI countries:
France, Germany, the Netherlands and the United Kingdon) and with the
participation of ISAS and NASA

\end{document}